\documentclass[12pt,preprint]{./aastex52/aastex}
\usepackage{natbib}
\usepackage{lscape}
\begin{document}
\title{Four new T dwarfs identified in PanSTARRS\,1 commissioning  data}
\shorttitle{Four T dwarfs in PanSTARRS\,1}
\author{Niall R. Deacon}
\affil{Institute for Astronomy, University of Hawai`i, 2680 Woodlawn Drive, Honolulu, HI 96822, USA}
\email{ndeacon@ifa.hawaii.edu}
\author{Michael C. Liu}
\affil{Institute for Astronomy, University of Hawai`i, 2680 Woodlawn Drive, Honolulu, HI 96822, USA}
\affil{Visiting Astronomer at the Infrared Telescope Facility, which is operated by the University of Hawaii under Cooperative Agreement no. NNX-08AE38A with the National Aeronautics and Space Administration, Science Mission Directorate, Planetary Astronomy Program}
\author{Eugene A. Magnier}
\affil{Institute for Astronomy, University of Hawai`i, 2680 Woodlawn Drive, Honolulu, HI 96822, USA}
\author{Brendan P. Bowler}
\affil{Institute for Astronomy, University of Hawai`i, 2680 Woodlawn Drive, Honolulu, HI 96822, USA}
\author{Bertrand Goldman}
\affil{Max Planck Institute for Astronomy, Koenigstuhl 17, D-69117 Heidelberg, Germany}
\author{Joshua A. Redstone}
\affil{Facebook, 1601 S. California Avenue, Palo Alto, CA 94304, USA}
\author{W. S. Burgett}
\affil{Institute for Astronomy, University of Hawai`i, 2680 Woodlawn Drive, Honolulu, HI 96822, USA}
\author{K. C. Chambers}
\affil{Institute for Astronomy, University of Hawai`i, 2680 Woodlawn Drive, Honolulu, HI 96822, USA}
\author{H. Flewelling}
\affil{Institute for Astronomy, University of Hawai`i, 2680 Woodlawn Drive, Honolulu, HI 96822, USA}
\author{N. Kaiser}
\affil{Institute for Astronomy, University of Hawai`i, 2680 Woodlawn Drive, Honolulu, HI 96822, USA}
\author{R.H. Lupton}
\affil{Princeton University Observatory, 4 Ivy Lane, Peyton Hall, Princeton University, Princeton, NJ 08544, USA}
\author{J.S. Morgan}
\affil{Institute for Astronomy, University of Hawai`i, 2680 Woodlawn Drive, Honolulu, HI 96822, USA}
\author{P.A. Price}
\affil{Princeton University Observatory, 4 Ivy Lane, Peyton Hall, Princeton University, Princeton, NJ 08544, USA}
\author{W.E. Sweeney}
\affil{Institute for Astronomy, University of Hawai`i, 2680 Woodlawn Drive, Honolulu, HI 96822, USA}
\author{J.L. Tonry}
\affil{Institute for Astronomy, University of Hawai`i, 2680 Woodlawn Drive, Honolulu, HI 96822, USA}
\author{R.J. Wainscoat}
\affil{Institute for Astronomy, University of Hawai`i, 2680 Woodlawn Drive, Honolulu, HI 96822, USA}
\author{C. Waters}
\affil{Institute for Astronomy, University of Hawai`i, 2680 Woodlawn Drive, Honolulu, HI 96822, USA}
 \label{firstpage}
 \begin{abstract}
A complete well-defined sample of ultracool dwarfs is one of the key 
science programs of the Pan-STARRS 1 optical survey telescope (PS1). Here 
we combine PS1 commissioning data with 2MASS to conduct a proper motion 
search (0.1--2.0\arcsec/yr) for nearby T dwarfs, using optical+near-IR 
colors to select objects for spectroscopic followup. The addition of 
sensitive far-red optical imaging from PS1 enables discovery of nearby 
ultracool dwarfs that cannot be identified from 2MASS data alone.  
We have searched 3700 sq. deg. of PS1 y-band (0.95--1.03 um) data to 
y$\approx$19.5 mag (AB) and J$\approx$16.5 mag (Vega) and discovered four 
previously unknown bright T dwarfs. Three of the objects (with spectral types T1.5, T2 and 
T3.5) have photometric distances within 25 pc and were missed by previous 
2MASS searches due to more restrictive color selection criteria. The 
fourth object (spectral type T4.5) is more distant than 25 pc and is only a single-band 
detection in 2MASS. We also examine the potential for completing the 
census of nearby ultracool objects with the PS1 3$\pi$ survey.
 \end{abstract}
 \section{Introduction}
 The Panoramic Survey Telescope And Rapid Response System (Pan-STARRS, \citealt{Kaiser2002}) is a pioneering wide-field, multi-filter, multi-epoch astronomical survey program. The project is planned to consist of four two-meter class telescopes operating from the Hawaiian Islands. The first of these, the Pan-STARRS\,1 telescope (PS1), has recently began full science operations on May the 15th 2010 on Haleakal\={a} on Maui and is operated by the Pan-STARRS\,1 Science Consortium. The largest of the PS1 surveys is the 3$\pi$ Survey which is planned to cover the entire sky visible from Hawai`i (3$\pi$ steradians in area, $\delta > -30^{\circ}$) in five filters ($g$, $r$, $i$, $z$ and $y$) with pairs of observations in each filter being taken at six different epochs. This will allow the survey to serve a range of science goals by both stacking individual exposures for deep images and using multiple epochs to identify moving or variable objects. So far the data available have been used to search for Trans-Neptunian Objects (\citealt{Wang2009}) and supernovae (e.g. \citealt{Botticella2010}). One of the key science areas where PS1 aims to contribute is the study of the local low luminosity population. The unique combination of a wide field, multiple epochs in the $y$ band and the survey strategy is designed to allow parallax measurements of faint local ultracool dwarfs. This could yield a volume-limited sample of the coolest substellar objects unhindered by presumptive color selection.

Most substellar objects are cool enough that they lie beyond the M spectral type in the standard Morgan-Keenan classification system. These objects fall either into the L spectral class (\citealt{Kirkpatrick1999}) or the cooler still T dwarf class (\citealt{Burgasser2006}). T dwarfs show deep, wide molecular absorption bands caused by water and methane in their near-infrared spectrum (1$-$2.5$\mu$m). Combined with their cool temperature  
these bands cause T dwarfs to have redder optical and bluer infrared colors than the earlier L-type objects. The absorption bands cause a double-peaked spectral shape in the 1$-$1.3$\mu$m region (the $y$ and $J$ bands) while a change in shape and suppression of the spectrum in the $H$ and $K$ bands due to water and methane absorption and collision induced absorption from $H_2$ molecules is also seen (\citealt{Burgasser2006}). The spectra also show significant absorption lines from potassium. The fluxes in the methane and water near-infrared absorption bands are used in the standard T dwarf spectral classification scheme of \cite{Burgasser2006} which superseeds  the previous classification schemes of \cite{Burgasser2002} and \cite{Geballe2002}.

Searches for ultracool dwarfs in wide-field surveys often rely on color selection. Studies based on the 2 Micron All Sky Survey (2MASS; \citealt{Skrutskie2006}, work by \citealt{Burgasser2004} and references therein) use infrared color cuts to exclude earlier type objects that, while excellent for discovering mid-late T dwarfs, can exclude redder earlier T-type objects. While the near infrared proper motion survey of \cite{Kirkpatrick2010} (which utilises overlap regions in the 2MASS survey) has more relaxed color selection criteria, it covers less than 10\% of the sky. The Sloan Digital Sky Survey (SDSS; \citealt{York2000}) studies by \cite{Chiu2006} and references therein use red optical $i-z$ colors to select targets. This choice allows them to detect objects across the L/T boundary, but are insensitive to later type objects due to their reliance on optical photometry. The UKIRT Deep Sky Survey (UKIDSS, \citealt{Lawrence2007}) has been exploited for late-T dwarfs by \cite{Goldman2010} and \cite{Burningham2010} and references therein. These studies have pushed the boundaries of the coolest T dwarfs but exclude the earlier T dwarfs by color selection.

Here we outline the results of a search to identify new, bright T dwarfs in Pan-STARRS\,1 commissioning data.

\section{Candidate Selection}
In the run-up to full science operations (from June 2009 to March 2010), the PS1 telescope took data in an observing mode matching the full survey strategy in order to provide information on data quality and survey efficiency. This dataset has provided us with observations in the 0.95-1.03~\micron~$y$ band, with some areas having additional coverage in the $z$ band. The filter profile of the PS1 $z$ band is different to that used by SDSS. Hence they cannot be directly compared without taking into account a color term. However like SDSS, Pan-STARRS\,1 reports AB magnitudes. Faint ultracool objects discovered in wide-field surveys such as PS1 are relatively nearby, allowing us to filter distant background contaminants by requiring proper motion. PS1 $y$ band data are only single epoch or have a maximum time baseline of three months, but, by combining these with information from other surveys, we are able to measure proper motions. The ideal companion survey for our first epoch of Pan-STARRS\,1 data is 2MASS. This provides both a second epoch with a time baseline of approximately ten years and photometry in the near infrared $J$, $H$ and $K_s$ bands. Thus, these additional data allow us to refine our photometric selection as well as provide the opportunity for astrometric selection. Although 2MASS has been extensively mined for ultracool dwarfs either as a dataset in its own right (\citealt{Burgasser2004}) or in combination with other surveys (e.g. \citealt{Metchev2008}), the addition of Pan-STARRS\,1 photometry means we can explore faint 2MASS detections excluded by  previous studies which were often limited to J=15.8.
\subsection{PS1 colors of ultracool dwarfs}
\label{colors}
\begin{figure}[htbp]
\begin{center}
\epsscale{0.7}
\plotone{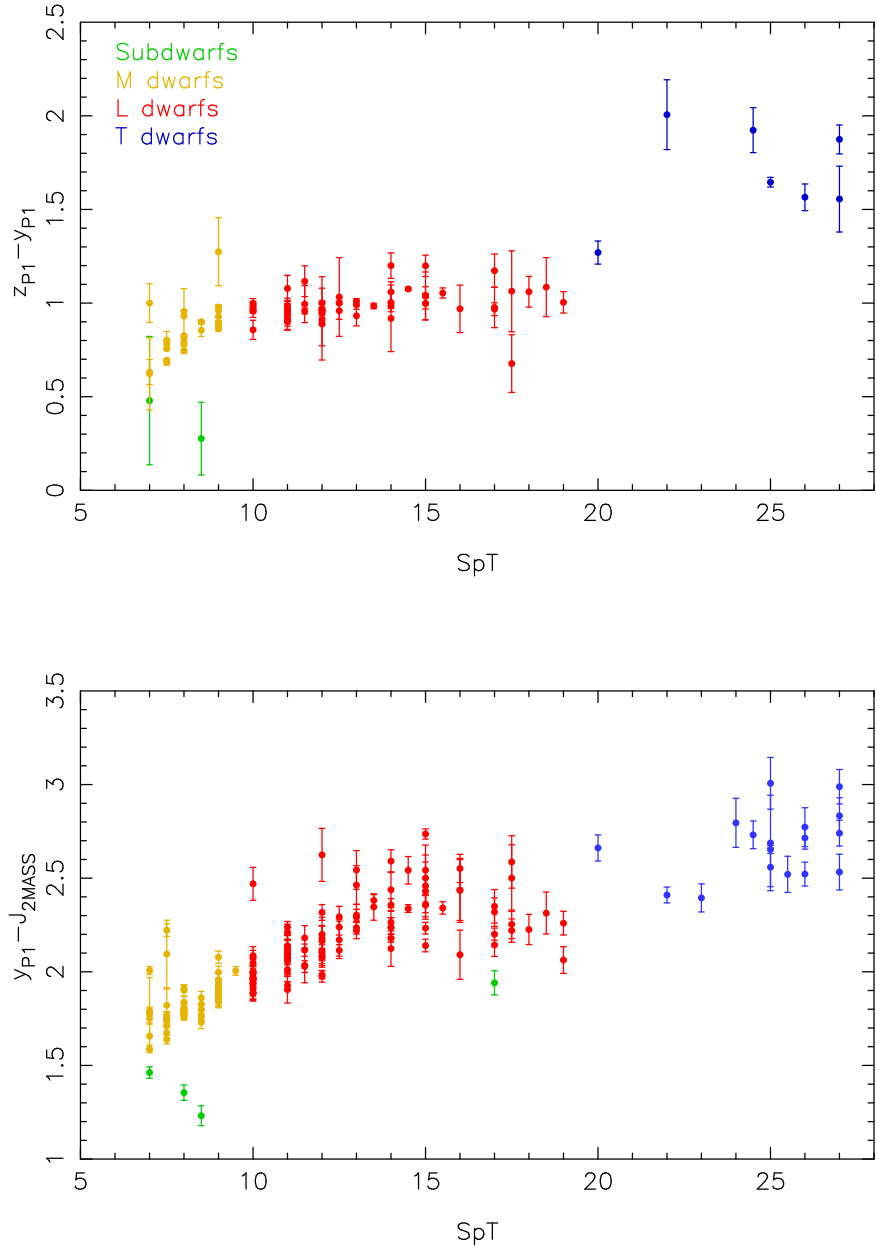}
\caption{Colors of ultracool dwarfs from \cite{Faherty2009} in the PS-1 and 2MASS filter systems. The green points are ultracool subdwarfs, the yellow points M dwarfs, red points L dwarfs and blue points T dwarfs. Note spectral type = 0 at M0, 10 at L0 and 20 at T0. The $z$ and $y$ bands are AB magnitudes. Spectral types are near-infrared for T dwarfs and optical otherwise.}
\label{fahertyplot}
\end{center}
\end{figure}

In order to determine appropriate color selection criteria to identify ultracool dwarfs we used the sample of late-type objects with astrometric solutions from \cite{Faherty2009} and corrected their positions to epoch 2010.0. We then searched the available PS1 data around these positions with a pairing radius of 5 arcseconds. These results were combined with the published 2MASS magnitudes to produce $z-y$ and $y-J$ colors (Figure~\ref{fahertyplot}). It is clear that ultracool dwarfs have distinctive red $z-y$ and $y-J$ colors and hence these can be used for candidate selection. It should be noted that PS1 magnitudes are from a preliminary version of the calibration system (\citealt{Magnier2006}, \citealt{Magnier2007}) which uses 2MASS photometry to calculate zero-points. Our internal consistency checks indicate these are in reasonable agreement with external datasets such as SDSS ( with a typical scatter due to systematic errors of 0.03mag in the $z$-band). 
\subsection{Color Cuts}
\label{Ccuts}
\begin{figure}[htbp]
\begin{center}
\epsscale{0.7}
\plotone{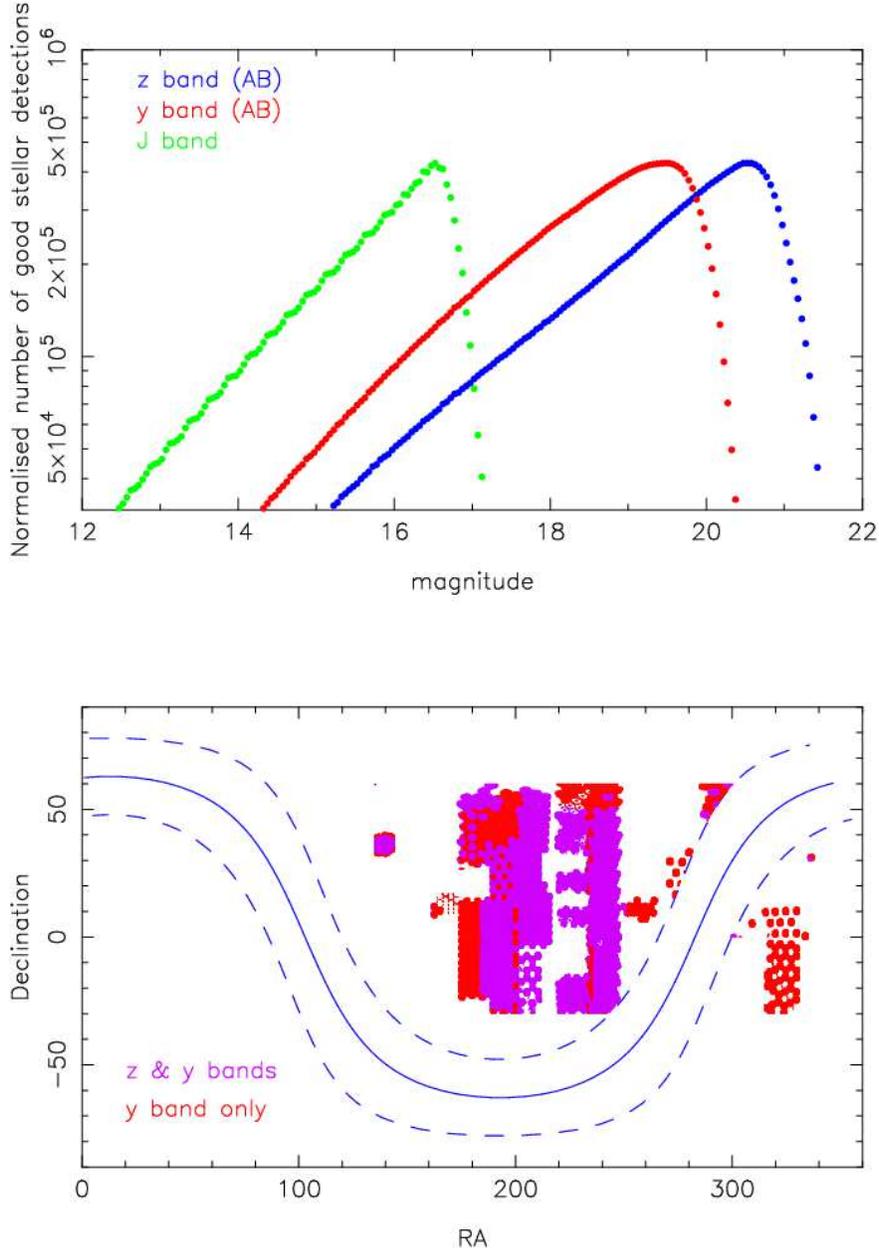}
\caption{The survey area covered by our commissioning data (red for $y$ band, purple for $z$ and $y$ band) along with magnitude histograms for different passbands ($z$ and $J$ histograms normalised to the height of the $y$ band histogram peak). For the PS1 data we required that the objects had good stellar 5$\sigma$ double detections to mimic our survey selection. Our survey begins to become incomplete at $J\sim16.5$ magnitudes, $y\sim19.5$ magnitudes and $z\sim20.5$ magnitudes. In the lower panel the solid blue line represents the Galactic Plane and the dashed blue line our $|b| > 15 ^{\circ}$ cut. These data were taken before the set-up of the PS1 telescope was finalized and so their quality may be lower than the final survey.}
\label{depthplot}
\end{center}
\end{figure}

We searched 3700 sq.deg. of PS1 $y$ band commissioning data taken before March 2010 with right ascensions between 9 hours and 24 hours. Of these data, 2500 sq.deg. had complementary $z$ band coverage (Figure~\ref{depthplot}). These coverages were calculated by summing the number of 0.1$\times$0.1 degree boxes which contained PS1 detections and then applying a fill factor of 74\% to take account of masked pixels around bright stars, chip gaps and bad cells. Candidates were identified as having good quality point source $y$ band detections on two or more images with the total significance of the mean of these detections $>5\sigma$ and no corresponding $J$ detection within one arcsecond in 2MASS. The Pan-STARRS\,1 survey strategy involves taking two images of the same area of sky approximately thirty minutes apart. Hence our double detection requirement will exclude transient sources such as asteroids and potential image artifacts. We then searched around the PS1 positions for objects which had a 2MASS $J$ band detection (not filtering on 2MASS photometry flags), no corresponding PS1 $z$ or $y$-detection within one arcsecond and with a maximum distance between the 2MASS and PS1 positions being the epoch difference between the two observations multiplied by our maximum proper motion of two arcseconds per year. When more than one 2MASS pair was found for a PS1 source we kept all matches which met our criteria. We selected candidates with $2.2<y-J<5.0$ and, where objects had PS1 $z$ band photometry, we required that $z-y>0.6$. These cuts were designed to exclude the bulk of low mass stars (spectral type M) from the sample. As we were searching for candidate T dwarfs, we also required a blue near-infrared color of $J-H<1.0$ (including quoted $H$ band upper limits). To reduce contamination from faint background stars in PS1 being paired with transient sources (such as uncataloged asteroids in 2MASS), all objects with a USNO (\citealt{Monet2003}) counterpart with $R$ or $I$ band magnitudes brighter than 20.5 and 18.5 respectively within 6" of either the PS1 or 2MASS positions were excluded. We undertook this step as true faint, red brown dwarfs are unlikely to have counterparts on relatively insensitive optical photographic plates. This initial query yielded 21,990 candidates. Additionally we searched both positions for each of our objects against the SuperCOSMOS Sky Survey (\citealt{Hambly2001}) with the same pairing radius and removed objects with $R$ or $I$ band detections brighter than 20.5 and 18.5 respectively and/or multiple band detections (i.e. objects with $B$ and $R$ detections, detections in both $R$ epochs or $R$ and $I$ detections). While SuperCOSMOS and USNO use mostly the same plate material, this method provides an additional check that objects do not have clear photographic plate counterparts. After the SuperCOSMOS cross-match stage fourteen objects with counterparts in the Dwarf Archives\footnote{http://dwarfarchives.org/} database of known brown dwarfs. The USNO $I$ band images of the remaining 16520 candidates were also visually inspected to ensure all objects with clear counterparts were excluded.  Following this final visual inspection stage, 474 objects remained. 

The previously known objects from Dwarf Archives which survived the SuperCOSMOS cross-match are detailed in Table~\ref{previous}. To test  the completeness of our survey, we cross-matched T dwarfs from \cite{Faherty2009} with the PS1 database searching for 5$\sigma$ $y$ double detections. In addition to our seven T dwarfs recovered, four are detected but have right ascensions earlier than nine hours so they did not fall in to our sample. Two fell outside our color cuts, one (SDSS J151114.66+060742.9) is inferred to be a mid L/mid T binary by \cite{Burgasser2010} and was marginally too red in $J-H$, while the other (SDSS J120602.51+281328.7) was too blue in $y-J$ by less than 0.02 magnitudes. Additionally one object was excluded due to a noisy ($\sigma>0.2$) $y$ band measurement, while three objects lay close to USNO or SuperCOSMOS detections. Additionally we extracted $y$ band images from the commissioning data for all T dwarfs in \cite{Faherty2009} with no y-band double detection. One object appeared in only one $y$ band detection. Four objects were in the footprints of multiple images but fell into masked regions such as chip gaps or poor cells. These four objects are 19\% of T dwarfs which fell into the footprint of two or more PS1 commissioning images. This number is in agreement with the 26\% loss or area estimated in the fill factor used to calculate our survey area.

\section{Follow-up Observations}
\subsection{Imaging}
We obtained follow-up imaging of candidates using WFCAM (\citealt{Casali2007}) on the UK Infrared Telescope (UKIRT). As our PS1 data are separated by ten years from 2MASS, many of our supposedly red high proper motion objects will in fact be anonymous blue Pan-STARRS\,1 objects paired with 2MASS transient sources such as uncataloged asteroids. Such PS1 sources will be too faint to appear in 2MASS, USNO-B or SuperCOSMOS. Additionally many of our sample are low signal to noise single $J$ band detections in 2MASS. While these could be the most scientifically interesting (due to their faint, blue infrared photometry), many are expected to be noise spikes or other image artefacts. Follow-up infrared imaging allows us to confirm that the candidates are as red in $y-J$ as the PS1 and 2MASS photometry suggest and to improve on the signal-to-noise of our 2MASS infrared photometry.

\begin{figure}[htbp]
\begin{center}
\epsscale{0.7}
\plotone{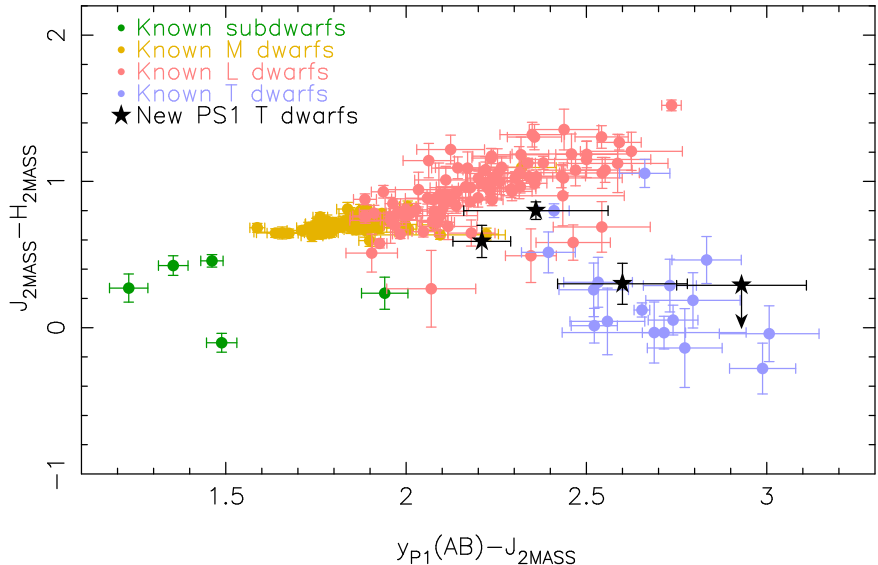}
\caption{Colors of ultracool dwarfs from \cite{Faherty2009} in the PS1 and 2MASS filter systems. The green points are ultracool subdwarfs, the yellow points M dwarfs, red points L dwarfs and blue points T dwarfs. Also plotted are our four spectroscopic targets. The arrow indicates a $J-H$ upper limit. The $y$ bands is in AB magnitudes.}
\label{fahertyplotcc}
\end{center}
\end{figure}

\begin{figure}[htbp]
\begin{center}
\epsscale{0.9}
\plotone{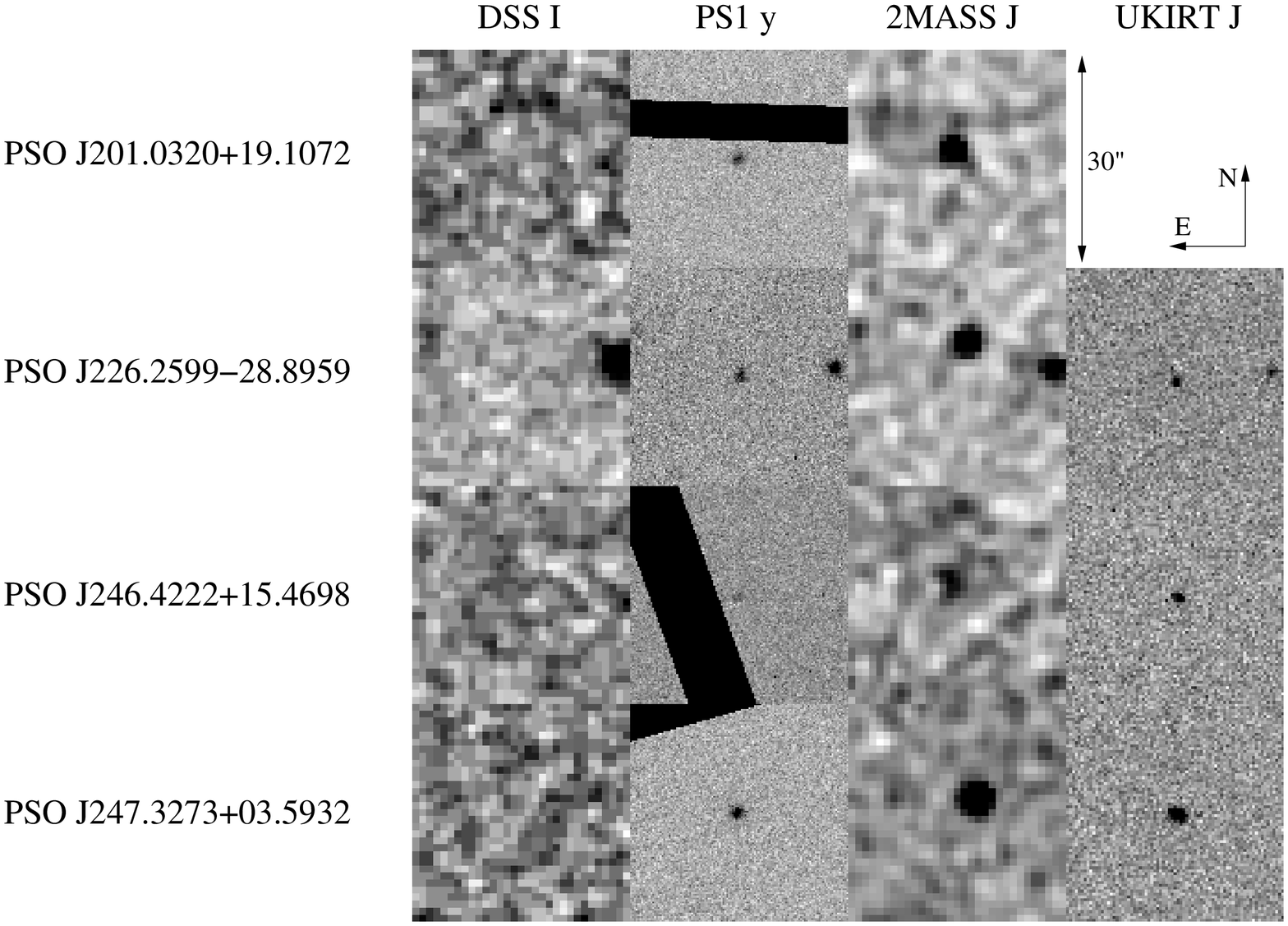}
\caption{\label{finders}Images for our four candidate objects centered on the PS1 position. The offsets in the position of the 2MASS detection are due to proper motion. The black streaks on the images are gaps between the chips of the PS1 camera. Note PSO~J201.0320+19.1072 was spectroscopically confirmed before our UKIRT observing run and hence has no UKIRT photometry.}
\end{center}
\end{figure}

Over the course of three and a half nights, a total of 195 candidates were imaged in $Y$, $J$, $H$ and $K$ on the 16-19 June 2010 (HST). Observations were  reduced at the Cambridge Astronomical Survey Unit using the WFCAM  survey pipeline \citep{Irwin2004, Hodgkin2009}. Additionally 56 objects had four band photometry in the UKIDSS Large Area Survey \citep{Lawrence2007}. As many of our candidates will in fact be spurious associations, we plotted the difference between the $J$ band magnitudes from 2MASS and WFCAM/UKIDSS. This produced a bimodal distribution with a division at 0.6 magnitudes. We chose this as the division between objects with good and discrepant photometry\footnote{While there is a color term between these two filters this will be $<0.4$ magnitudes \citep{Stephens2004}, smaller than our chosen acceptable $J$ band discrepancy of 0.6 magnitudes. We list the 193 objects with discrepant photometry in Table~\ref{duds1}.} We then selected objects which appeared to be red (i.e. $y_{P1}-J_{MKO}>2.0$). These were given further visual screening and had their photometry from the Sloan Digital Sky Survey (which was not used in our intial color cuts) examined. before the four best candidates were followed up spectroscopically. These candidates are plotted on a color-color diagram in Figure~\ref{fahertyplotcc} and images for the object are shown in Figure~\ref{finders}. The remaining objects which were judged not to have discrepant photometry and which we judged to be not interesting enough for spectroscopic follow-up were likely late M or L dwarfs which scattered into the T dwarf sample. These were not considered interesting enough for spectroscopic follow-up. 
\begin{figure}[htbp]
\begin{center}
\epsscale{0.9}
\plotone{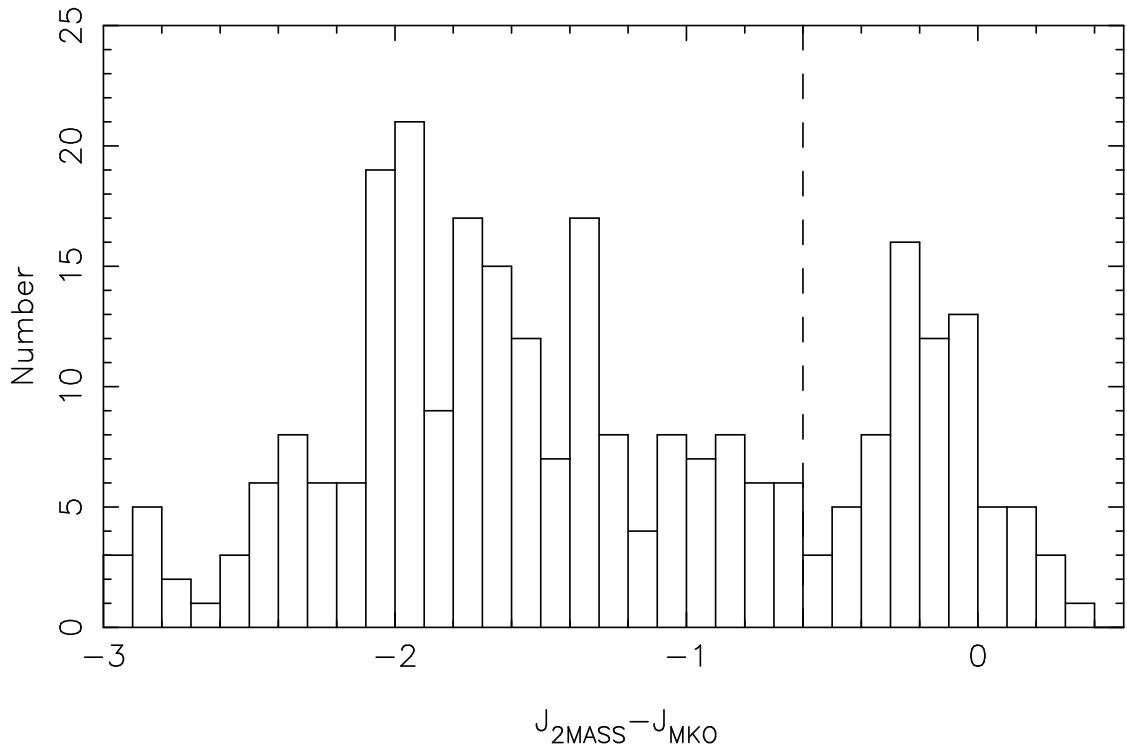}
\caption{\label{JJhist}The difference between the $J$ band magnitudes from the 2MASS survey and those measured from our UKIRT photometry. Note the bimodal distribution. We take the division between these two modes at 0.6 magnitudes (as shown by the dashed line) as the boundary between objects with discrepant and non-discrepant photometry.}
\end{center}
\end{figure}

\subsection{Spectroscopy}

We obtained low-resolution ($R\approx$75--120) spectra of objects selected
by their WFCAM and PS1 photometry from NASA's Infrared Telescope Facility (IRTF) located on
Mauna Kea, Hawai`i. Conditions were clear. We used the facility near-IR
spectrograph SpeX \citep{Rayner2003} in prism mode, obtaining
0.8--2.5~\micron\ spectra in a single order. We oriented the slit with the parallactic
angle to minimize the effect of atmospheric dispersion. Each target was
nodded along the slit in an ABBA pattern, with individual exposure times
of 120--150~sec, while the telescope was guided using the off-axis
optical guider. For flux and telluric calibration, we observed A0~V
stars contemporaneously with the science targets and at
comparable airmass and sky location. All spectra were reduced using
version~3.4 of the SpeXtool software package
\citep{Cushing2004, Vacca2003}. A summary of the
observations is provided in Table~\ref{spexobs}.

\subsection{Infrared spectral types}

\begin{figure}[htbp]
\begin{center}
\epsscale{0.9}
\plotone{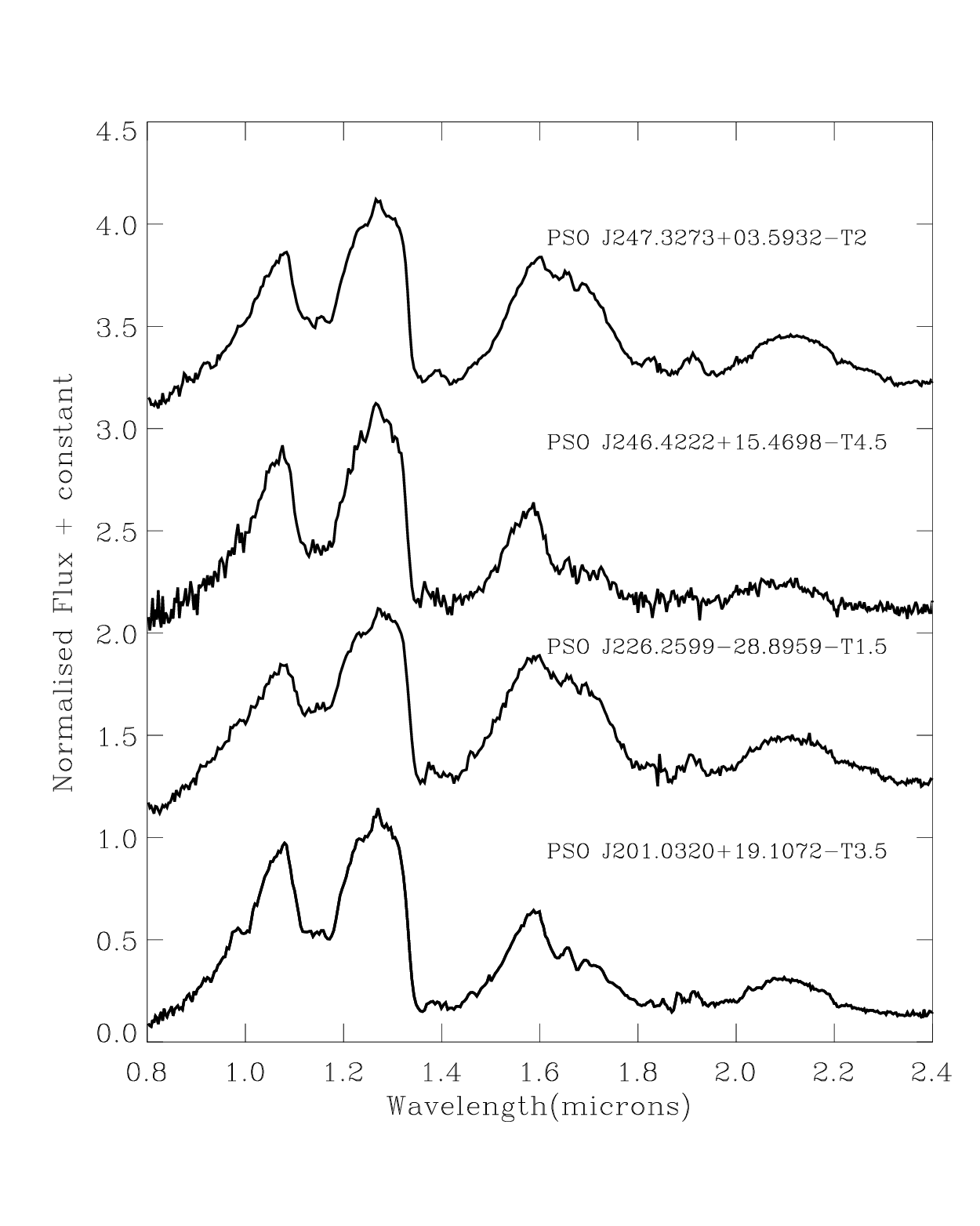}
\caption{\label{spectraplot}Spectra for our objects spectral types derived from the classification system of \cite{Burgasser2006}.}
\label{spectra}
\end{center}
\end{figure}

Figure~\ref{spectraplot} presents our near-IR spectra, showing the strong water and
methane absorption bands that are the hallmarks of the T~spectral class.
We assigned spectral types by measuring the five flux indices defined by
\citet{Burgasser2006} and then applying the polynomial relations
of \citet{Burgasser2007}.
We also visually determined spectral types by comparing with IRTF/Spex
prism spectra of the T~dwarfs spectral standards chosen by \cite{Burgasser2006}. For each object, the depth of the H$_2$O and CH$_4$ absorption
bands were examined, normalizing the spectra of the objects and the standards
to their peak fluxes in the $J$, $H$, and $K$ bands individually. 
The types from the flux indices agreed well with those from visual
examination. For PSO~J226.2599$-$28.8959, the two results disagreed by
1~subclass, therefore we took the average to assign the final spectral type
(Table~\ref{spexclass}).

\section{Discussion}
\subsection{Photometric distances}
We calculated photometric distance estimates for our objects based on spectral type vs. absolute magnitude relations. For PSO~J247.3273+03.5932\footnote{PSO (Pan-STARRS Object) is the standard IAU designation for objects detected in Pan-STARRS.}, PSO~J246.4222+15.4698 and PSO~J226.2599-28.8959 we used our UKIRT photometry and the relations of \cite{Liu2006}. This paper quotes both a bright relation which includes suspected and confirmed binaries and a faint relation which excludes these. The disagreement between these relations is most pronounced in the mid-Ts at approximately one magnitude. As the binarity of our sample of four objects is unknown, we use an average of the bright and faint relations to estimate the intrinsic brightness of our objects. For PSO~J201.0320+19.1072 we used 2MASS photometry and the relations of \cite{Liu2006} converted into the 2MASS system using the relations of \cite{Stephens2004}. The distance estimates for each filter were then averaged to produce the values shown in Table~\ref{objphotometry}\footnote{Note the PS1 positions quoted here and in other tables are derived from our astrometric solutions and calculated for an epoch of 2010.0}. Three of our objects PSO~J201.0320+19.1072, PSO~J247.3273+03.5932 and PSO~J226.2599-28.8959 have distance estimates within 25 pc with PSO~J246.4222+15.4698 lying further away at 31.3 pc.

\subsection{Our objects in other surveys}
Our four objects are relatively bright and thus could have been detected by previous work such as studies of the 2MASS survey (\citealt{Burgasser2004} and references therein) the Sloan Digital Sky Survey (SDSS; \citealt{York2000}, \citealt{Chiu2006}, and references therein) and the UKIRT Deep Sky Survey (UKIDSS, \citealt{Lawrence2007}, \citealt{Deacon2009a}, \citealt{Goldman2010},\citealt{Burningham2010} and references therein). The Burgasser 2MASS searches used two different $H-K_s$ cuts and made photometric selections from 2MASS data products before the final calibration. For comparison we use the final bluer cut $H-K_s<0.0$ and photometry from the final 2MASS catalog.

\noindent {\bf PSO~J201.0320+19.1072}: The UKIDSS observation for this object is from Data Release 7 while the most recent publications (\citealt{Goldman2010}, \citealt{Burningham2010}) only include data up to Data Release 4. It is too red in $H$-$K_s$ (0.06) for the 2MASS T dwarfs surveys. However, it should appear in  \cite{Chiu2006} due to its red SDSS colors. We are unsure as to why it does not appear in this study as its observation date of May 2005 is later than most objects in Chiu et al's catalog, but one object appears from June 2005. 

\noindent {\bf PSO~J226.2599-28.8959}: This object falls outside the UKIDSS and SDSS survey areas and is too red ($H$-$K_s$=0.30) to be detected by 2MASS searches for T dwarfs. 

\noindent {\bf PSO~J246.4222+15.4698}: This is similarly not in UKIDSS and is too faint in $J$ (16.77) to fall into the 2MASS studies' selection sample. It appears in the SDSS database that would have been searched by  \cite{Chiu2006}, but has a high error on the $z$ magnitude, which could have lead to it falling outside their sample.

\noindent {\bf PSO~J247.3273+03.5932}: This object is not in UKIDSS or SDSS and is too red in $H$-$K_s$ (0.39) to appear in the 2MASS T dwarf samples.

\begin{figure}[htbp]
\begin{center}
\epsscale{0.7}
\plotone{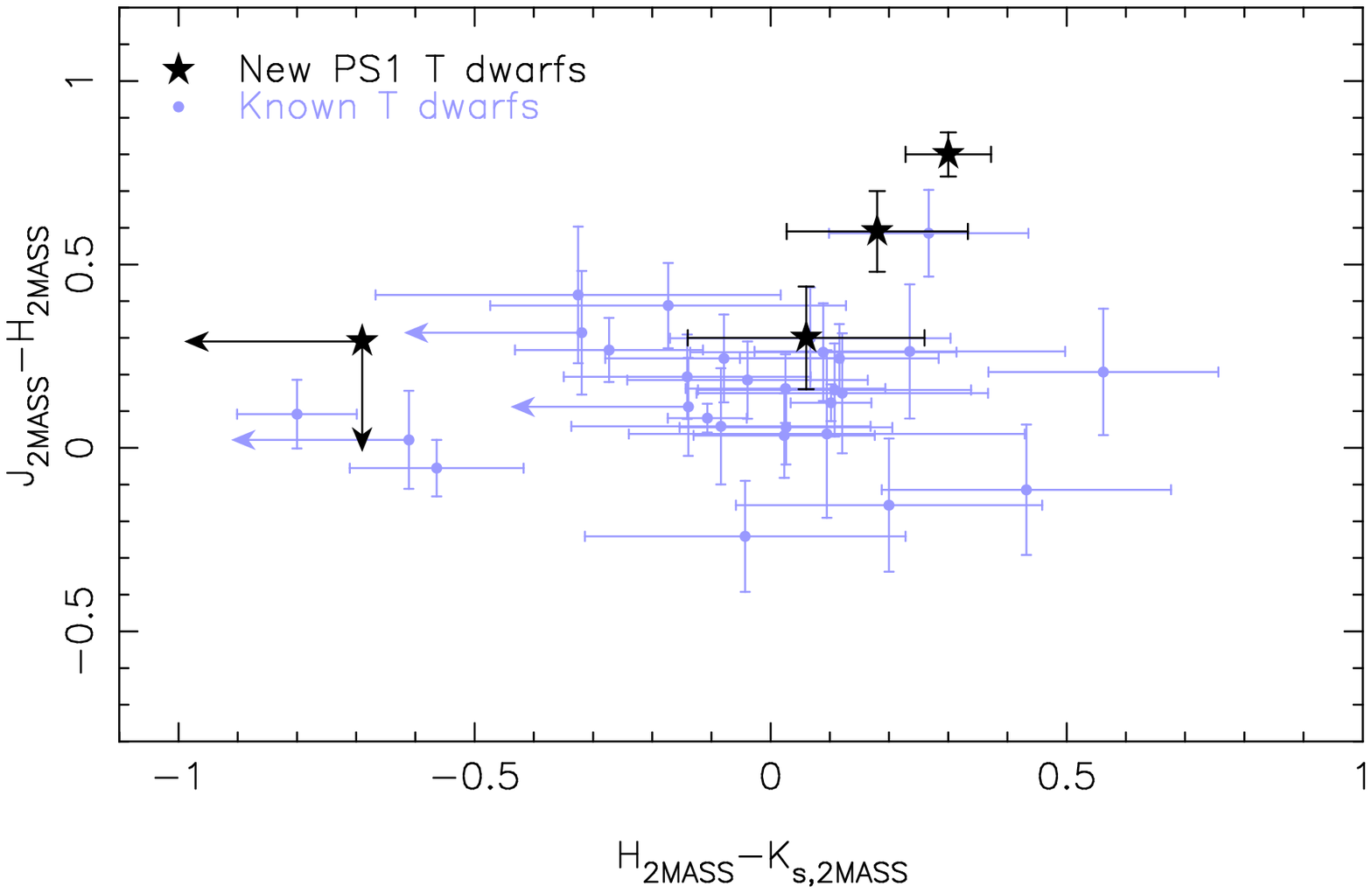}
\caption{Our four new T dwarfs plotted against discoveries from \cite{Burgasser1999}, \cite{Burgasser2000}, \cite{Burgasser2000b}, \cite{Burgasser2003b} and \cite{Burgasser2004}. Three of our objects appear redder than the bulk of T dwarfs discovered in these 2MASS searches. The colors of the Burgasser et al. objects spread beyond their color selection criteria ($J-H<0.3$, $H-K_s<0.0$). This is due to their selection being made before the final 2MASS calibration and some objects having colors from the final calibration which take them outside the initial selection.}
\label{burgasserplot}
\end{center}
\end{figure}

Three of our objects appear redder than the final color cuts of the Burgasser searches for T dwarfs. Figure~\ref{burgasserplot} shows how our objects compare to the discovered in these earlier studies.

\subsection{Prospects for future Pan-STARRS\,1 ultracool dwarf searches}
\begin{figure}[htbp]
\begin{center}
\epsscale{0.7}
\plotone{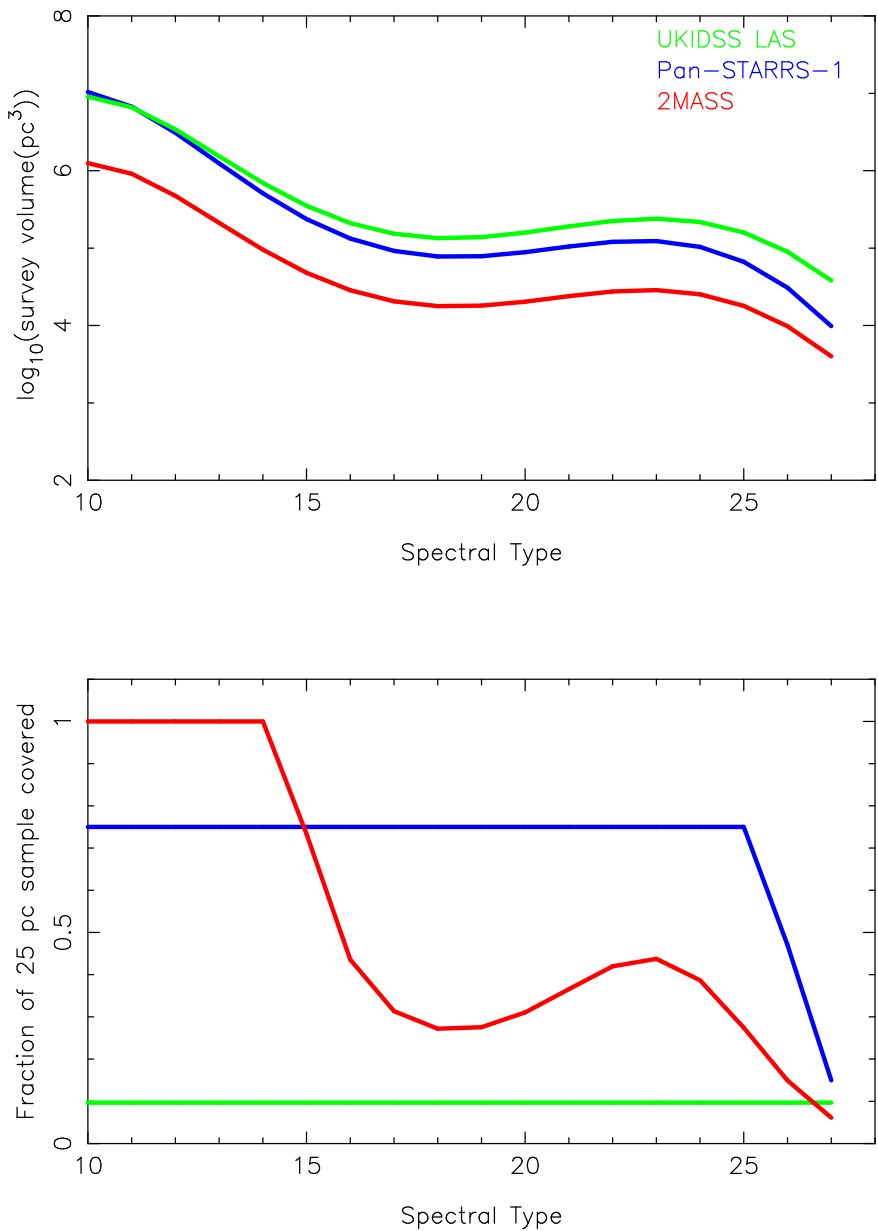}
\caption{Top panel: The expected volume in cubic parsecs for the UKIDSS LAS, Pan-STARRS\,1 3$\pi$ survey and 2MASS survey for ultracool dwarfs. It is clear that PS1 will outperform 2MASS and will be competitive with UKIDSS down to the mid-T dwarfs. Spectral Type=0 at M0, 10 at L0 and 20 at T0. Lower panel: The fraction of the 25 pc. sample covered by the three surveys. PS1 can only access three quarters of the sky and is hence limited to 75\% for most spectral types but will provide a complete sample within this area down to the mid-T regime. See text for assumptions used.}
\label{volumeplot}
\end{center}
\end{figure}

To characterize the parameter space covered by the Pan-STARRS\,1\,3$\pi$ survey and to compare with other leading wide-field surveys for ultracool dwarfs we calculated the expected volume each survey would cover for different spectral types. For the UKIDSS survey we averaged the bright and faint spectral type - absolute magnitude relations of \cite{Liu2006} and assumed a $J$ band completeness limit of 18.8 (\citealt{Burningham2010}). In the case of 2MASS most searches to date have only gone to the nominal completeness limit of $J=15.8$, so this was used along with the absolute magnitude relations of \cite{Liu2006} transformed into the 2MASS filter system using the relations from \cite{Stephens2004}. Searches for T dwarfs based on 2MASS are less complete than this estimate as these objects are blue and the 2MASS $H$ and $K_s$ bands are shallower than the $J$ band. Conversely 2MASS will be deeper in $H$ and $K_s$ for the reddest L dwarfs. For PS1 we also used the \cite{Liu2006} and \cite{Stephens2004} relations and added an empirical spectral type vs. $y-J$ relation derived from the data in Figure~\ref{fahertyplot}. For our $y$ band limit we used 19.5 (see Figure~\ref{depthplot}) and assume a future PS1 only search\footnote{\cite{Magnier2008} and \cite{Beaumont2010} outline how PS1 can be used for $y$ band parallax only studies of the local brown dwarf population.}. The results can be seen in Figure~\ref{volumeplot}. It is clear that in absolute volume terms PS1 is competitive with UKIDSS until about T4 and will perform better than 2MASS across the full range of spectral types plotted. Furthermore, UKIDSS is a deeper, narrower survey, so objects discovered in PS1 will be brighter and easier to follow up. This is well demonstrated by the lower panel of Figure~\ref{volumeplot} which shows that Pan-STARRS\,1 will provide a reliably complete survey of the solar neighbourhood into the mid-T regime.
\section{Conclusions}
We have identified the first new T dwarfs from the Pan-STARRS\,1 3$\pi$ survey. This survey will likely lead to the most complete sample of brown dwarfs in the solar neighbourhood. Of our four discoveries, three were bright enough to be detected by previous surveys, but were excluded due to the restrictive color selection. This shows that Pan-STARRS\,1 based surveys will be able to loosen color selection criteria, detect substellar extrema missed by work in the field so far and provide the most complete survey to date of brown dwarfs in the solar neighborhood. 
\acknowledgements
The PS1 Surveys have been made possible through contributions of the Institute for Astronomy, the University of Hawaii, the Pan-STARRS Project Office, the Max-Planck Society and its participating institutes, the Max Planck Institute for Astronomy, Heidelberg and the Max Planck Institute for Extraterrestrial Physics, Garching, The Johns Hopkins University, the University of Durham, the University of Edinburgh, Queen's University Belfast, the Harvard-Smithsonian Center for Astrophysics, and the Los Cumbres Observatory Global Telescope Network, Incorporated, the National Central University of Taiwan, and the National Aeronautics and Space Administration under Grant No. NNX08AR22G issued through the Planetary Science Division of the NASA Science Mission Directorate.The United Kingdom Infrared Telescope is operated by the Joint Astronomy Centre on behalf of the Science and Technology Facilities Council of the U.K. 
This paper makes use of observations processed by the Cambridge Astronomy
Survey Unit (CASU) at the Institute of Astronomy, University of Cambridge.
 The authors would like to thank Mike Irwin and the team at CASU for making the reduced WFCAM data available promptly, Tim Carroll and Watson Varricatt for assisting with UKIRT observations and Will Best for helping to prepare the target list for these. They would also like to thank John Rayner and Alan Tokunaga for making engineering time available for IRTF observations and Paul Sears, Dave Griep, Bill Golisch and Eric Volqardsen for assisting with these. This research has benefitted from the SpeX Prism Spectral Libraries, maintained by Adam Burgasser at http://www.browndwarfs.org/spexprism. This publication makes use of data products from the Two Micron All Sky Survey, which is a joint project of the University of Massachusetts and the Infrared Processing and Analysis Center/California Institute of Technology, funded by the National Aeronautics and Space Administration and the National Science Foundation. This research has benefitted from the M, L, and T dwarf compendium housed at DwarfArchives.org and maintained by Chris Gelino, Davy Kirkpatrick, and Adam Burgasser. E.A.M. and M.L. were supported by NSF grant AST 0709460; E.A.M. was
also supported by AFRL Cooperative Agreement FA9451-06-2-0338. This project was supported by DFG-Sonderforschungsbereich 881 "The Milky Way System".\\
{\it Facilities:} \facility{IRTF (SpeX)}, \facility{PS1}, \facility{UKIRT}
\bibliography{ndeacon}
\bibliographystyle{apj}
\begin{table}
\caption{Previously known objects found in PS1 commissioning data. Citation key:  a - \cite{Knapp2004}, b - \cite{Fan2000}, c - \cite{Chiu2006},  d - \cite{Leggett2000}, e - \cite{Hawley2002}, f - \cite{Burgasser2000}, g - \cite{Burgasser2002}, h - \cite{Strauss1999} \label{previous}}
\scriptsize
\begin{tabular}{llccccccccccr}
\hline
Name&SpT&$z$&$\sigma_z$&$y$&$\sigma_y$&$J$&$\sigma_J$&$H$&$\sigma_H$&$K_S$&$\sigma_K$&Reference\\
\hline
  SDSS J115553.86+055957.5 & L7.5 & - & - & 18.00 & 0.02 & 15.66 & 0.08 & 14.7 & 0.07 & 14.12 & 0.07 & a\\
  SDSSp J120358.19+001550.3 & L3 & - & - & 16.25 & 0.02 & 14.01 & 0.02 & 13.06 & 0.02 & 12.48 & 0.02 & b\\
  SDSS J125011.65+392553.9 & T4 & - & - & 19.17 & 0.12 & 16.54 & 0.11 & 16.18 & 0.18 & 16.06 & 0.25 & c\\
  SDSSp J125453.90-012247.4 & T2$^5$ & 19.31 & 0.19 & 17.30 & 0.01 & 14.89 & 0.03 & 14.09 & 0.02 & 13.84 & 0.05 & d\\
  SDSS J134525.57+521634.0 & L3.5 & 20.61 & 0.08 & 19.75 & 0.09 & 16.94 & 0.18 & 16.26 & 0.24 & 15.35 & 0.14 & c\\
  SDSS J135852.68+374711.9 & T4.5 & 21.09 & 0.11 & 19.16 & 0.04 & 16.46 & 0.09 & 16.14 & 0.17 & 16.1 & 0.0 & c\\
  SDSS J140231.75+014830.3 & L1 & 18.37 & 0.02 & 18.15 & 0.04 & 15.45 & 0.06 & 14.65 & 0.07 & 14.18 & 0.07 & e\\
  Gliese 570D & T8 & 19.93 & 0.06 & 18.06 & 0.05 & 15.32 & 0.05 & 15.27 & 0.09 & 15.24 & 0.16 & f\\
  SDSS J154508.93+355527.3 & L7.5 & 20.38 & 0.09 & 19.71 & 0.13 & 16.83 & 0.17 & 16.0 & 0.19 & 15.43 & 0.16 & c\\
  2MASSI J1553022+153236 & T7 & 20.20 & 0.16 & 18.65 & 0.07 & 15.82 & 0.07 & 15.94 & 0.16 & 15.51 & 0.18 & g\\
  SDSS J162255.27+115924.1 & L6 & 20.09 & 0.09 & 19.12 & 0.08 & 16.88 & 0.18 & 16.15 & 0.22 & 15.55 & 0.2 & c\\
  SDSSp J162414.37+002915.6 & T6$^5$ & 19.77 & 0.06 & 18.21 & 0.03 & 15.49 & 0.05 & 15.52 & 0.10 & 15.52 & 0.0 & h\\
  SDSS J162838.77+230821.1 & T7 & - & - & 19.40 & 0.08 & 16.46 & 0.10 & 16.11 & 0.15 & 15.87 & 0.24 & c\\
  SDSS J214046.55+011259.7 & L3 & - & - & 18.35 & 0.07 & 15.89 & 0.08 & 15.31 & 0.09 & 14.42 & 0.08 & e\\
\hline
\end{tabular}
\end{table}
\clearpage
\begin{table}
\caption{Objects with discrepant photometry between 2MASS and UKIRT/UKIDSS. For these purposes we define discrepant as objects whose $J$ band photometry differs by more than 0.6 magnitudes between the two measurements. All coordinates are J2000. This is an example of the format used in the full version of this table which is available electronically.  \label{duds1}}
\scriptsize
\begin{tabular}{llrrllrrrrrr}
\hline
RA (2MASS)&Dec (2MASS)&$J_{2MASS}$&$\sigma_J$&RA (PS1)&Dec (PS1)&$y_{P1}$&$\sigma_y$&$Y_{MKO}$&$\sigma_Y$&$J_{MKO}$&$\sigma_J$\\
\hline
  11:35:49.89 & +04:33:18.3 & 16.56 & 0.15 & 11:35:49.06 & +04:33:27.4 & 19.28 & 0.12 & 18.40 & 0.03 & 17.81 & 0.04\\
  11:38:34.63 & +04:41:04.7 & 16.42 & 0.15 & 11:38:33.81 & +04:41:16.0 & 20.01 & 0.12 & 19.27 & 0.06 & 18.72 & 0.07\\
  11:39:32.50 & +00:49:05.6 & 16.50 & 0.14 & 11:39:33.10 & +00:48:49.1 & 20.14 & 0.14 & 19.86 & 0.14 & 19.00 & 0.12\\
  11:43:01.81 & +10:53:41.3 & 16.47 & 0.15 & 11:43:01.90 & +10:53:38.8 & 20.32 & 0.14 & 19.67 & 0.10 & 19.01 & 0.10\\
  11:48:05.09 & +09:28:08.8 & 16.63 & 0.16 & 11:48:05.19 & +09:28:01.1 & 20.04 & 0.12 & 19.66 & 0.07 & 19.05 & 0.09\\
  11:55:50.23 & +03:29:30.9 & 16.42 & 0.15 & 11:55:49.18 & +03:29:39.4 & 19.97 & 0.12 & 18.99 & 0.05 & 18.47 & 0.06\\
  11:56:10.24 & +05:14:04.9 & 16.53 & 0.13 & 11:56:10.40 & +05:14:10.6 & 19.45 & 0.08 & 18.59 & 0.04 & 18.06 & 0.04\\
  12:01:10.66 & -00:09:05.4 & 16.82 & 0.15 & 12:01:10.46 & -00:09:14.9 & 19.72 & 0.09 & 18.97 & 0.07 & 18.36 & 0.11\\
  12:02:09.47 & -00:19:03.8 & 17.01 & 0.15 & 12:02:10.13 & -00:18:54.3 & 19.57 & 0.19 & 18.73 & 0.06 & 18.14 & 0.06\\
\hline
\end{tabular}
\end{table}
\normalsize
\begin{table}
\caption{\label{spexobs} Spectroscopic follow-up using SpeX on IRTF.}
\footnotesize
\begin{tabular}{lccccc}
\hline
Object&UT date&Airmass&Slit&T$_{int}$&Spectral\\
&&&arcseconds&s&Resolution\\
\hline
PSO~J201.0320+19.1072&2010-05-17&1.06&0.8$\times$15&1200&75\\
PSO~J226.2599-28.8959&2010-07-15&1.72&0.8$\times$15&960&75\\
PSO~J246.4222+15.4698&2010-06-19&1.77&0.5$\times$15&1440&120\\
PSO~J247.3273+03.5932&2010-06-19&1.50&0.5$\times$15&720&120\\
\hline
\end{tabular}
\normalsize
\end{table}
\begin{table}
\caption{\label{spexclass} The spectral indices for the four T dwarfs found in this study.}
\scriptsize
\begin{tabular}{lcccccccc}
\hline
Object&H2O-J&CH4-J&H2O-H&CH4-H&CH4-K&avg/RMS&Visual&Final\\
\hline
PSO~J201.0320+19.1072&0.437 (T3.2) & 0.551 (T2.4) & 0.413 (T4.0) & 0.631 (T3.6) & 0.349 (T3.7) & T3.4$\pm$0.6 &  T3.5 &  T3.5 \\
PSO~J226.2599-28.8959 &0.539 (T1.5) & 0.696 (L8.8) & 0.551 (T1.3) & 0.871 (T1.8) & 0.670 (T1.2) & T0.9$\pm$1.2 &  T2   &  T1.5 \\
PSO~J246.4222+15.4698&0.310 (T4.5) & 0.493 (T3.6) & 0.366 (T4.7) & 0.448 (T4.9) & 0.272 (T4.5) & T4.5$\pm$0.5 &  T5   &  T4.5 \\
PSO~J247.3273+03.5932&0.435 (T3.2) & 0.626 (T0.7) & 0.446 (T3.5) & 0.925 (T1.4) & 0.608 (T1.8) & T2.1$\pm$1.2 &  T2   &  T2   \\
\hline
\end{tabular}
\normalsize
\end{table}
\begin{landscape}
\begin{table}
\caption{\label{objphotometry}Photometry and astrometry for our four T dwarfs.} \tablecomments{Values with errors marked U are 2MASS 95\% confidence upper limits. Proper motions are in arcseconds per year. Photometric distances are calculated using the relations of \cite{Liu2006} (for MKO) and \cite{Liu2006} and \cite{Stephens2004} (for 2MASS) with errors estimated using the typical RMS scatter quoted for the absolute magnitude relations (0.4 magnitudes). UKIRT photometry is in the MKO system. Note PSO~J201.0320+19.1072 was confirmed spectroscopically prior to our UKIRT run and thus has no MKO photometry. The two objects with SDSS photometry have only their $z$ band magnitudes listed. As previously stated the SDSS and PS1 $z$ bands have different filter profiles, hence the two are not directly comparable. All other bands for both were fainter than 23rd magnitude. Our astrometry errors have 2 contributions: a systematic floor of ~80
mas from the 2MASS reference frame, and shot noise of ~100 - 200 mas
for these faint detections \citep{Magnier2008}.  With a baseline of roughly 10 years, our
proper motion accuracy is approximately 20-30 milliarcseconds per
year. The positions in the object designations are derived from our astrometric solutions and calculated to an epoch of 2010.0.}
\vspace{2mm}
\tiny
\begin{tabular}{lccccccccccc}
\hline
name&SpT&$\mu_{\alpha}\cos(\delta)$&$\mu_{\delta}$&$z$&$y$&$Y$&$J$&$H$&$K/K_s$&d$_{phot}$&Source\\
\hline
PSO~J201.0320+19.1072&T3.5&-0.10&-0.09&-&18.37$\pm$0.17&&15.77$\pm$0.07&15.47$\pm$0.12&15.41$\pm$0.16&20$\pm$4 pc&PS1/2MASS\\
&&&&19.00$\pm$0.04&&&&&&&SDSS\\
PSO~J226.2599-28.8959&T1.5&0.10&-0.44&19.45$\pm$0.06&18.04$\pm$0.04&&15.83$\pm$0.07&15.24$\pm$0.08&15.06$\pm$0.13&&PS1/2MASS\\
&&&&&&16.95$\pm$0.03&15.86$\pm$0.02&15.29$\pm$0.02&14.9$\pm$0.02&21$\pm$4 pc&UKIRT\\	
PSO~J246.4222+15.4698&T4.5&-0.24&-0.24&-&19.7$\pm$0.10&&16.77$\pm$0.15&16.48 U&17.17 U&&PS1/2MASS\\
&&&&20.13$\pm$0.17&&17.97$\pm$0.05&16.77$\pm$0.03&16.91$\pm$0.04&17.14$\pm$0.10&34$\pm$7 pc&SDSS/UKIRT\\
PSO~J247.3273+03.5932&T2&0.24&-0.16&19.11$\pm$0.04&17.65$\pm$0.02&&15.29$\pm$0.04&14.48$\pm$0.04&14.18$\pm$0.06&&PS1/2MASS\\
&&&&&16.19$\pm$0.02&15.10$\pm$0.01&14.53$\pm$0.01&14.28$\pm$0.02&15$\pm$3 pc&UKIRT\\
\hline
\end{tabular}
\normalsize
\end{table}
\clearpage

\normalsize

\end{landscape}
\clearpage
\end{document}